\documentclass[preprint,showpacs,amsmath,amssymb]{revtex4}

    \usepackage{graphicx}
    \usepackage{dcolumn}
    \usepackage{bm}
    \usepackage{color}
    \newcommand{\VEC}[1]{\mbox{\boldmath${#1}$}}

\begin{document}

\title{
  Wide cylindrical or spherical optical potentials from laser beam superpositions}

\author{Ole Steuernagel}
\affiliation{School of Physics, Astronomy and Mathematics, University of
Hertfordshire, Hatfield, AL10 9AB, UK }

\email{O.Steuernagel@herts.ac.uk}

\date{\today}

\begin{abstract}
  Superpositions of paraxial Laguerre-Gauss laser beam modes to generate
  optical potentials based on the optical dipole force are
  investigated theoretically. Parabolic and other monomial potentials with
  even powers, in cylindrical and spherical symmetry,
  with large diameters, can be synthesized.  This
  superposition approach promises to help with high quality atom cloud
  manipulation and imaging.
\end{abstract}

\pacs{
32.80.Lg, 
32.80.Qk, 
42.50.Vk, 
42.60.Jf  
}

\maketitle

\section{Introduction}

The fields of atom~\cite{Meschede_Metcalf_03} and molecular
optics~\cite{Seideman_JChPh97,Stapelfeldt_PRL97} have developed
considerably over recent years. Atomic beams that behave like laser
beams can be created~\cite{BlochPRL99}; they are considered for
direct deposition~\cite{Meschede_Metcalf_03}, used in
microscopy~\cite{NotteAIPC_07}, precision
metrology~\cite{Marion_PRL03}, and for the studies of quantum
statistical effects~\cite{Greiner_NAT02,Greiner_NAT03} such as the
Hanbury Brown-Twiss effect~\cite{Jeltes_NAT07}.

The quality of the preparation of atomic ensembles, particularly in
the cases of ultra-cold gases, is excellent, frequently as good as
fundamental limits allow~\cite{Bloch_RMP08,Jeppesen_PRA08}. The same
is not true of atom-optical components, for example atom-beam
focussing~\cite{Meschede_Metcalf_03} suffers from the lack of lenses
with large numerical apertures, see~\cite{Ole_PRA09} and references
therein. Similarly, wide parabolic optical dipole potentials for
focussing and defocussing~\cite{Jeltes_NAT07} of atomic clouds at
tolerable laser beam power~\cite{Miossec_OC02,Ole_PRA09} are
currently unavailable.

Here, the use of the optical dipole force for the application of
wide high-quality aberration-free pulsed or stationary 2D and 3D
isotropic potentials, primarily for the manipulation of cold gas
clouds, is considered. In the regime of laser light far detuned from
the atomic transition the resulting potentials are conservative and
their strength is proportional to the laser light
intensity~\cite{Metcalf.book,Meschede_Metcalf_03}. It is shown that
superpositions of Laguerre-Gaussian beams using spatial light
modulators allow us to tailor laser beams in a suitable
fashion~\cite{Ole_JOPA05} to create wide cylindrical and spherical
optical potentials improving control for the manipulation of atomic
or molecular clouds. Harmonic potentials are considered in greatest
detail because of their importance for focussing~\cite{Ole_PRA09}
and defocussing~\cite{Ole_PRA09,Jeltes_NAT07} in atom optics and
atomic lithography~\cite{Meschede_Metcalf_03} or for feeding into
waveguides~\cite{Miossec_OC02} of atomic chips~\cite{Fortagh_SCI05}.
The superposition approach can be extended to non-harmonic
potentials with cylindrical and spherical symmetry, see
section~\ref{S_Cylindrical_Potentials} below. Non-harmonic
potentials should facilitate dynamical studies of trapped atom
clouds~\cite{Murray_JPhB05}, perturbations, the formation of
caustics, and aspects of quantum-classical
behaviour~\cite{Friedman_PRL01}.

For focussing and defocussing of atomic beams many approaches have
been investigated, see references in~\cite{Ole_PRA09}. For focussing
and defocussing of cold gas clouds in two and three dimensions
magnetic
focussing~\cite{Cornell_PRL91,Hinds99,Knyazchyan_JPhCS05,Arnold_NJP06}
setups have been experimentally implemented~\cite{Miossec_OC02}.
When miniaturizing such approaches care has to be taken that
unwanted interactions with bulk media do not disturb trapped
gas~\cite{Hohenester_Scheel_PRA07}. The superposition approach
introduced here is meant to complement these and make the great
flexibility of optical dipole force approaches more accessible.

In section~\ref{sec_LG_beams} we briefly recap the physics of
Laguerre-Gauss modes $LG_{p,l}$ and the optical dipole force. We
find that superpositions of members of the family~$\{ LG_{p,1},
p=0,\ldots,N_{max} \}$ are required for the generation of parabolic
potentials, other values of orbital angular momentum~$l$ can yield
purely monomial potentials of order~$2 l$. We determine the
expansion coefficients needed to create the desired optical
potentials with cylindrical symmetry in
section~\ref{S_Cylindrical_Potentials}. An important consequence of
the superposition approach and the main motivation for this work is
the ability to very considerably save laser power, see
subsection~\ref{SS_PowerSavings}. Wide aberration-free optical
dipole force potentials become realizable which currently are
infeasible because of laser beam dilution, this has been recognized
as a road block for some time: ``In spite of numerous impressive
achievements, using laser light interaction to tailor atom beams
demands a high quality of the transverse collimation and this
technique is difficult to scale with accessible laser
powers''~\cite{Miossec_OC02}.

The use of Laguerre-Gauss beams is not only experimentally well
established~\cite{Maurer_RitschMarte_NJP07} they also form an
appropriate basis for our analysis: limitations of the superposition
approach arise because of mode-dispersion due to Gouy's
phase~\cite{Ole_2005AmJPh,Ole_PRA09}. This phase is incorporated in
the definition of Laguerre-Gauss modes, see
Eq.~(\ref{eq_LG_modes_one_b}); its effects are considered in
subsections~\ref{SS_CYLINDRICAL_Mode_Dispersion}
and~\ref{SS_spherical_Mode_Dispersion}. We finally show how crossing
two modulated beams with cylindrical symmetry can be used to form a
spherically symmetrical potential in
section~\ref{S_spherical_Potentials} and conclude in
section~\ref{S_Conclusion}.

\section{Laguerre-Gauss Beams}\label{sec_LG_beams}

Laguerre-Gauss modes are monochromatic, paraxial beam
solutions~\cite{Ole_2005AmJPh,Allen_PRA92,Siegman.book,Haus.book,Pampaloni_2004}
and are defined as

\begin{eqnarray}
LG \left( p,l,z_R,\lambda_L,\rho,z \right) & = & \sqrt {{\frac {
2}{(1+\delta_{0l}) \pi} \frac{ p!}{ \left( p+l \right) !}}} \left(
{\frac {\sqrt {2}\; \rho}{w(z) }} \right)^{l} L \left( p,l,2\,{
\frac {{\rho}^{2}}{ w(z)^2 }} \right)
\nonumber \\
& \times &  \left( \frac{e^{i l\xi}}{w(z)} \right) {e^{-{\frac
{{\rho}^{2}}{
 w(z)^2 }}}}{e^{-i ( 2\,p+l+1 ) \phi(z) }} e^{ \frac{i k_L \rho^2}{2 R(z)} } \; .
\label{eq_LG_modes_one_b}
\end{eqnarray}

Here, the monochromatic plane wave factor $\exp[i(k_L z - \omega_L
t)]$ is
omitted~\cite{Ole_2005AmJPh,Allen_PRA92,Siegman.book,Haus.book,Pampaloni_2004},
$L$ are generalized Laguerre polynomials, $p$ and $l$ the integer
valued nodal, and angular momentum numbers respectively, $\VEC r =
(\VEC \rho,z)$ is the position vector with the transverse coordinate
vector~$\VEC \rho = (x,y)$; the Kronecker-delta function
$\delta_{0l}$ reflects the fact that the modes with zero angular
momentum have to be normalized differently to the other modes. The
transverse coordinates also parameterize the orbital angular
momentum phase~$\xi$ via the
relationship~$e^{i\xi}=x+iy$~\cite{Allen_PRA92}. The frequency of
the monochromatic laser~$\omega_L$ gives rise to its wavenumber~$k_L
= \omega_L/c = 2 \pi / \lambda_L$ where $\lambda_L$ is the laser
light's wavelength. The wave front radii $R(z)= (z^2 + z_R^2)/z$,
the beam radii $w(z)=w_0 \sqrt{ 1+z^2/z_R^2}$, with the focal beam
radius $w_0=\sqrt{\lambda_L z_R / \pi}$, and the longitudinal
Gouy-phase shifts $\phi(z)=\arctan(z/z_R)$ are all parameterized by
the beams' Rayleigh lengths
$z_R$~\cite{Siegman.book,Haus.book,Allen_PRA92,Pampaloni_2004}.

Correctly chosen superpositions of modes using spatial light
modulators~\cite{OleJMO05,Ole_JOPA05,Ole_PRA09}
\begin{equation}
{\Psi}_{P}(\VEC r) = \sum_{p=1}^{P} c_{Pp} \cdot LG \left(
p-1,1,z_R,\lambda_L,\rho,z \right)
 \label{eq_Mode_Sum}
\end{equation}
allows us to create superpositions~$\Psi$ which give rise to
parabolic intensity distributions. Following
reference~\cite{Haus.book} the use of ($y$-) polarized modes in
Equation~(\ref{eq_Mode_Sum})) yields an electric field which is
polarized in the $y$-direction with a small contribution in the
$z$-direction due to the tilt of wave fronts off the beam axis
($\hat{\bf x}, \hat{\bf y}, \hat{\bf z}$ are the unit-vectors and
$\Re$ stands for real-part)

\begin{eqnarray}
{\VEC E}_{P}({\VEC r};t) = \Re \{ [ \hat{\bf y} \; \omega_L \;
\Psi_{P} + \hat{\bf z} \; ic \; \frac{\partial \Psi_{P}}{\partial x}
] e^{i(k_L z - \omega_L t)} \} \; .
\label{true.E.field}
\end{eqnarray}

In keeping with the paraxial approximation of not overly focussed
beams we neglect the transverse derivative in
eq.~(\ref{true.E.field}). The associated time-averaged light
intensity distribution then has the form~\cite{Haus.book}

\begin{eqnarray}
I_{P}({\VEC r}) = \epsilon_0 \; \left\langle {\VEC E}_{P}({\VEC
r},t)^2\right\rangle \approx \frac{\epsilon_0}{2} \; \omega_L^2 \; |
\Psi_{P}({\VEC r}) |^2 \; . \label{Intensity}
\end{eqnarray}

{\subsection{Normalization, Intensity
    Scaling and Gradient Reduction}\label{SS_Norm_Intensity_singleMode}}

With the normalized modes of Eq.~(\ref{eq_LG_modes_one_b}) and
assuming that the sum of the coefficients $\sum |c_{p}|^2$ in
Eq.~(\ref{eq_Mode_Sum}) is normalized to unity we use the
cross-sectional beam power normalization

\begin{eqnarray}
\int_{-\infty}^\infty \int_{-\infty}^\infty dx \, dy \,
|\Psi_{P}(x,y,z)|^2 =\frac{2}{\epsilon_0\omega_L^2}
\int_{-\infty}^\infty \int_{-\infty}^\infty dx \, dy \, I_{P}(x,y,z)
\doteq \frac{2}{\epsilon_0\omega_L^2} \bar I_{P} =1 \,.
\label{Eq_Intensity_normalization}
\end{eqnarray}

We note that the intensity ${I(x,y,z)}$ of beams of fixed total
power reduces inversely proportionally to their width $w_0$ in one
direction, that is, their field amplitudes scale with $w_0^{-1/2}$in
$x$ and $y$. Furthermore the field gradients diminish with
$w_0^{-1}$. This implies that the effective curvature of the
integrated laser light intensity~$\frac{1}{2}\int_0^R |\nabla
\Psi|^2 2 \pi r dr$, responsible for atomic focussing, scales with
$w_0^{-4}$. We face an unfavourable quartic scaling with the beam
width if we attempt to expand a laser beam transversally in $\rho$
in order to widen the effective potential without weakening its
power. Additionally, as we will show below, pure modes have small
useful areas to generate the desired potentials, the combination of
these two factors makes a pure mode approach
unfeasible~\cite{Ole_PRA09,Miossec_OC02,Gallatin_1991OSAJB}.  It
forces us to employ the mode superpositions studied here. Below, we
discuss two approaches for the compensation of gradient weakening:
by power compensation in subsection~\ref{SS_Power_Compensation}, and
by beam waist narrowing in
subsection~\ref{SS_Rayleigh_Length_Matching}.

\subsection{Optical Dipole Force}\label{ss_Atoms}

We assume that the interaction between atoms and the laser light is
well described by a two-level scheme (excited state $e$ and ground
state $g$) in rotating wave approximation with effective atomic line
width~$\Gamma$ and resonance frequency~$\omega=\omega_e-\omega_g$.
This leads to the expression $ I(\VEC r )\,\Gamma^2/(2 I_S) =
\Omega(\VEC r )^2$ for the Rabi-frequency~$\Omega$ as a function of
the ratio of the local laser intensity~$I(\VEC r )$ and the
transition's saturation
intensity $I_S 
=\pi h c \Gamma/(3\lambda^3)$~\cite{Natarajan96,Metcalf.book}. With
sufficiently weak laser intensity~$I$ and sufficiently large
detuning~$\delta_\omega=\omega_L-\omega$ of the laser
frequency~$\omega_L$ from the atomic transition
frequency~$\omega$, the AC-Stark shift gives rise to a conservative
optical dipole potential which, to first order in~$I/I_S$, has the
form~\cite{Metcalf.book,Hope_PRA96}
\begin{eqnarray}
  U_\omega \approx \frac{\hbar}{8} \frac{ \Gamma^2}{\delta_\omega}
  \frac{I(\VEC r)}{I_S} \, \approx \frac{\epsilon_0 \hbar}{16}
  \frac{\Gamma^2 \omega_L^2}{\delta_\omega I_S} \; |\Psi_{P}({\VEC r}) |^2.
\label{dipole_potential_approx}
\end{eqnarray}

This potential is modified due to detrimental spontaneous emission
noise and light fluctuations. These tend to increase with increasing
laser intensity but can be decreased by increased
detuning~\cite{Metcalf.book} or through the use of more complicated
optical level schemes~\cite{Hope_PRA96}. Further discussion of their
influences is beyond the scope of this paper.

\subsection{Potentials of order $2l$}\label{SS_potential_order_2l}

We now consider cylindrical atom-potentials with purely monomial
modulation in the transverse direction~$\propto \rho^{2l}$. A
Taylor-expansion in $\rho$ shows that $LG$-modes with angular
orbital momentum $l$ depend in leading order on~$\rho^l$ and every
other higher order~($\rho^{l+2n}, n$ a positive integer). The Taylor
coefficients of different modes are linearly independent of each
other. Combining them into suitable superpositions created from mode
families~$\{ LG_{p,l}, p=0,\ldots,N_{max} \}$ allows us to retain
the leading and remove all higher order terms up to and including
that of order~$\rho^{l+2N_{max}}$. Almost purely monomial potentials
of order~$2 l$ therefore arise from such superpositions.

\section{Cylindrical Potentials}\label{S_Cylindrical_Potentials}

The coefficients for superpositions using the mode family~$LG_{p,1}$
are straightforward to determine through Gauss elimination. The
first six superpositions~$\{\Psi_P, P=1,\ldots,6\}$ yield the
following (normalized $\sum_{p=1}^{P} |c_{Pp}|^2 =1$) coefficient
matrix
\begin{eqnarray}
 \left[ c_{Pp} \right] =  \left[ \begin {array}{cccccc}
  1.0& 0.0& 0.0& 0.0& 0.0& 0.0\\\noalign{ } 0.9427&- 0.3333& 0.0& 0.0& 0.0& 0.0
\\\noalign{ } 0.8339&- 0.5360& 0.1313& 0.0& 0.0& 0.0
\\\noalign{ } 0.7212&- 0.6277& 0.2883&- 0.05546& 0.0& 0.0
\\\noalign{ } 0.6209&- 0.6467& 0.4150&- 0.1525& 0.02435& 0.0
\\\noalign{ } 0.5368&- 0.6261& 0.4959&- 0.2595& 0.08003&-
 0.01095\end {array} \right] \; .
\label{eq_coeff_matrix}
\end{eqnarray}

\begin{figure}[h]
\centering
\includegraphics[width=3.in,height=1.5in]{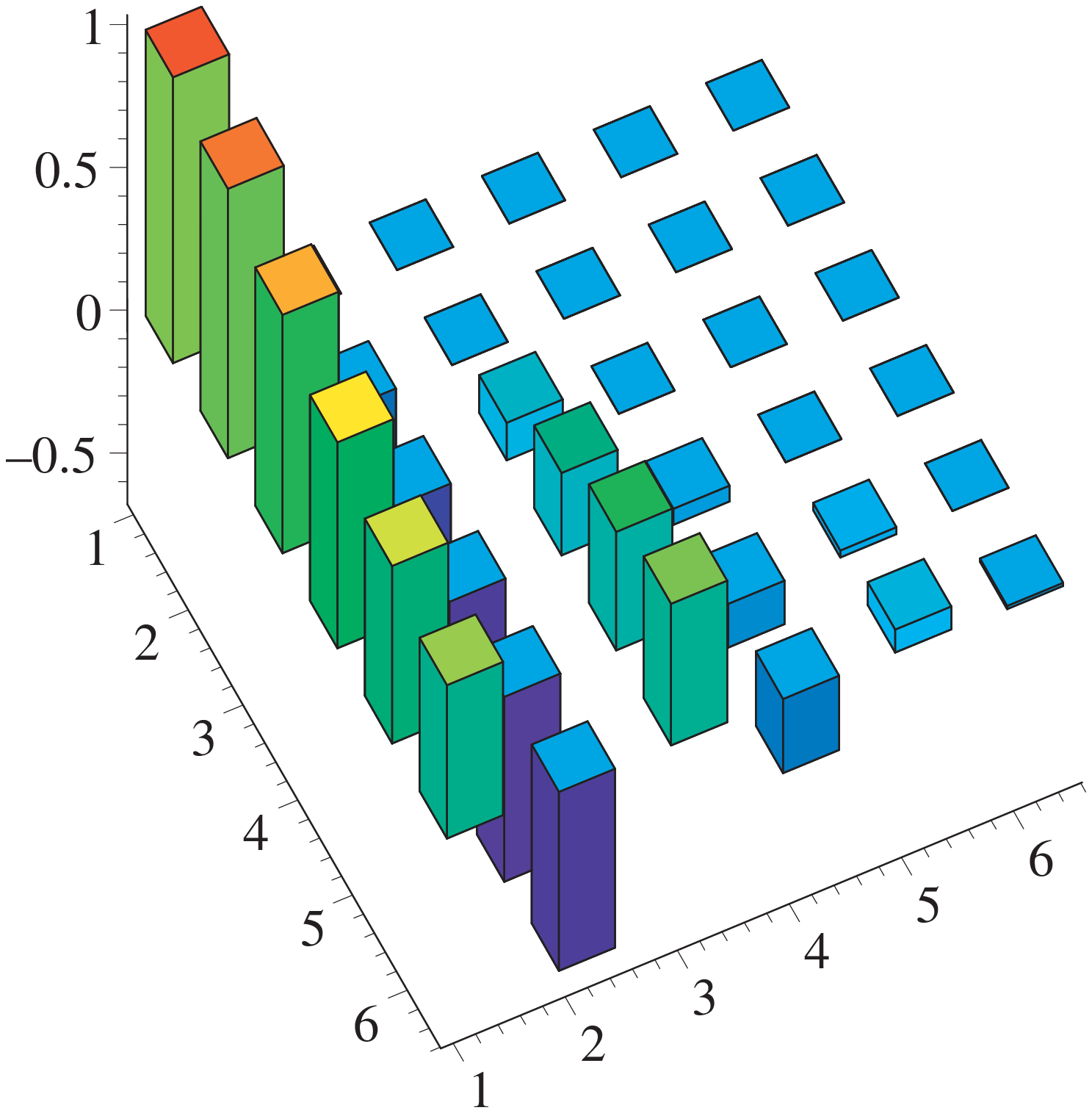}
\put(-220,110){\rotatebox{0}{\mbox{$c_{Pp}$}}} \put(-180,20){$P$}
\put(-20,10){$p$}
\hspace{0.9cm}
\includegraphics[width=3.in,height=1.5in]{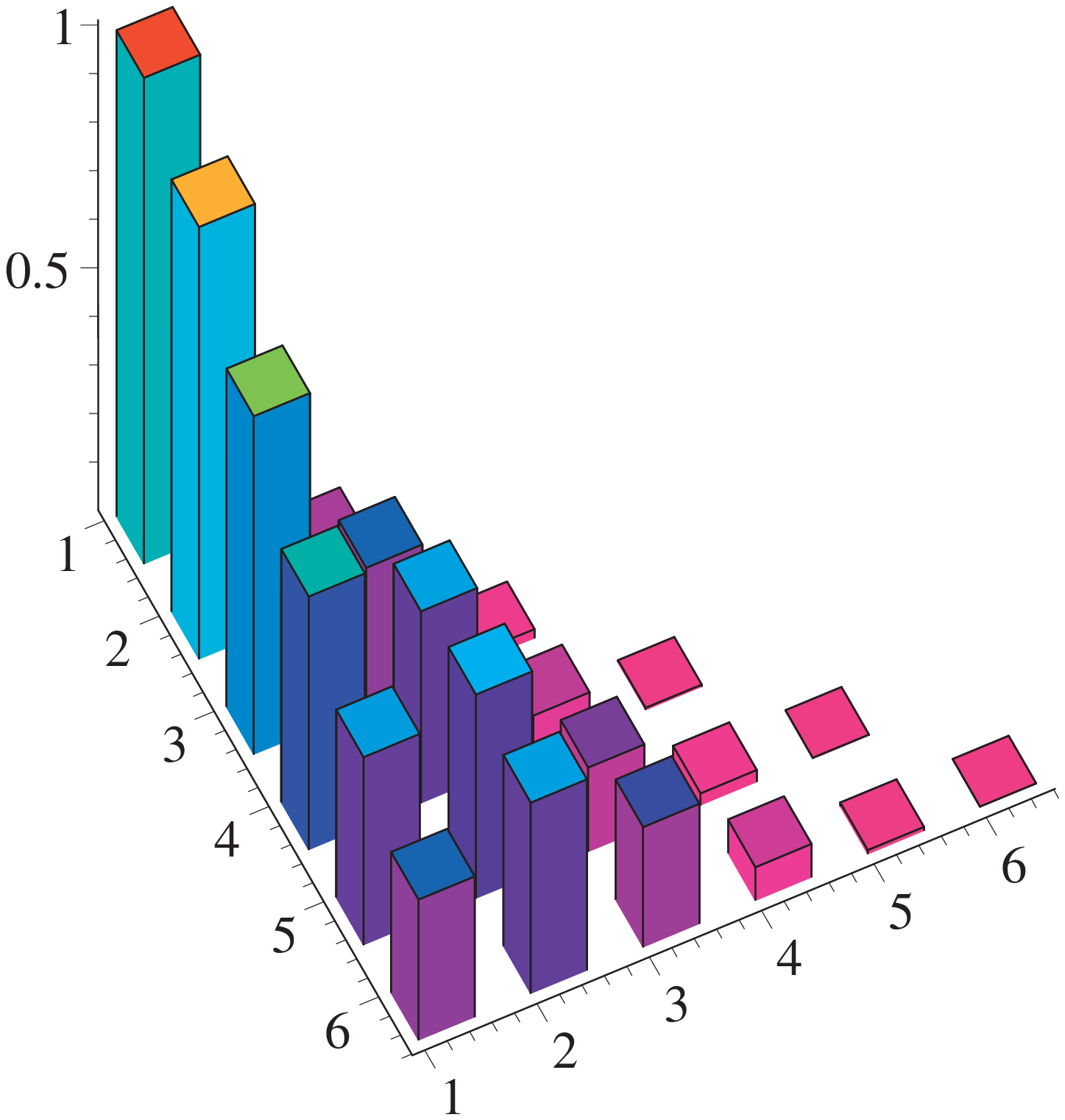}
\put(-230,100){\rotatebox{0}{\mbox{$c_{Pp}^2$}}} \put(-190,20){$P$}
\put(-20,10){$p$}
\caption{Amplitude coefficients~$c_{Pp}$ and
probabilities~$c_{Pp}^2$ of
  Laguerre-Gauss superpositions~$\Psi_{P}$ of up to
  $N_{max}=5{}^{\mbox{th}}$ order modes ($P = N_{max}+1=1,2,\ldots,6$)
  according to the coefficient matrix~(\ref{eq_coeff_matrix}).}
\label{Fig_coeff_matrix}
\end{figure}

\begin{figure}[h]
\centering
\includegraphics[width=3.in,height=2.5in]{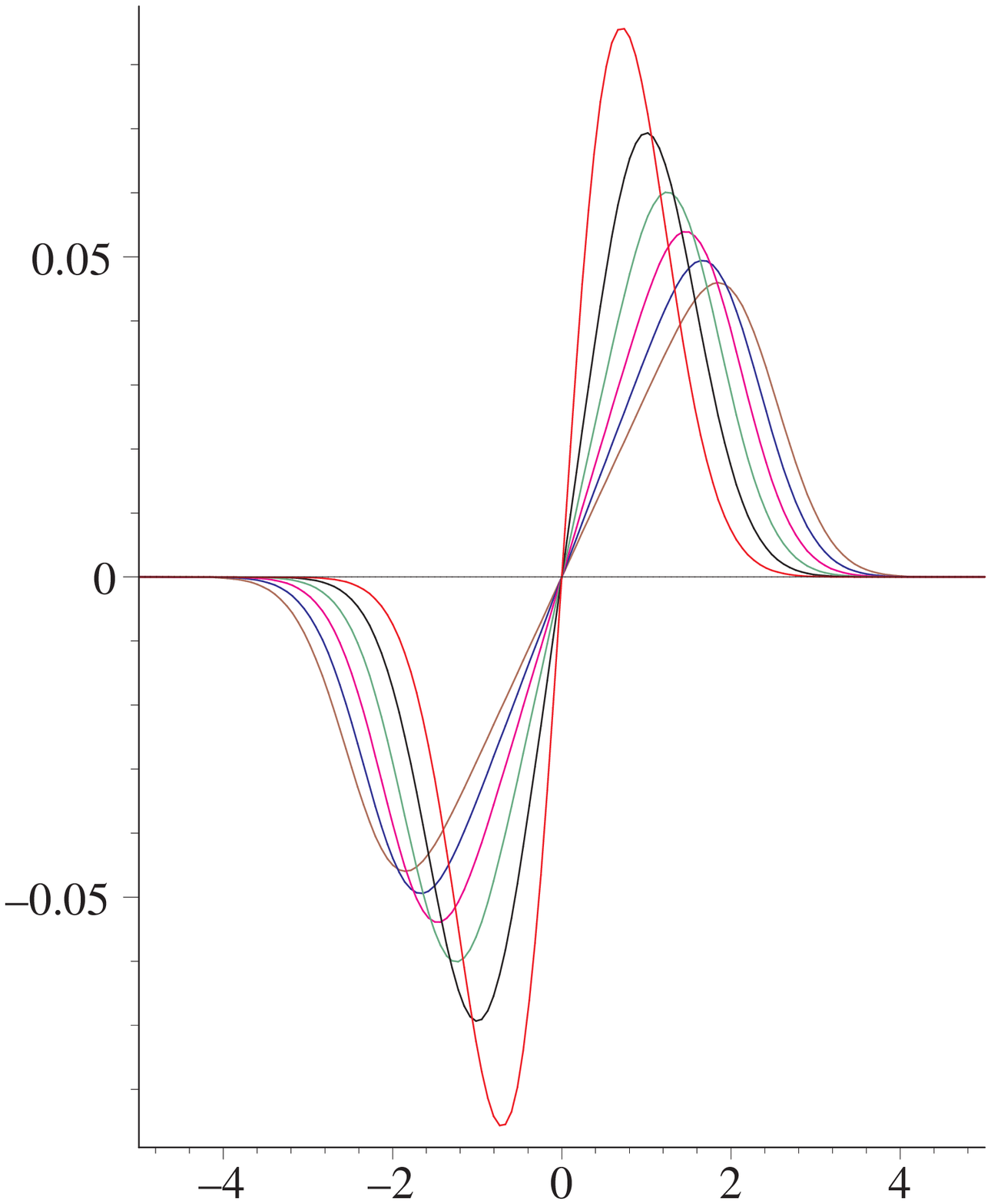}
\put(-210,160){\rotatebox{0}{\mbox{$E$}}} \put(-30,-10){$x[w_0]$}
\hspace{0.9cm}
\includegraphics[width=3.in,height=2.5in]{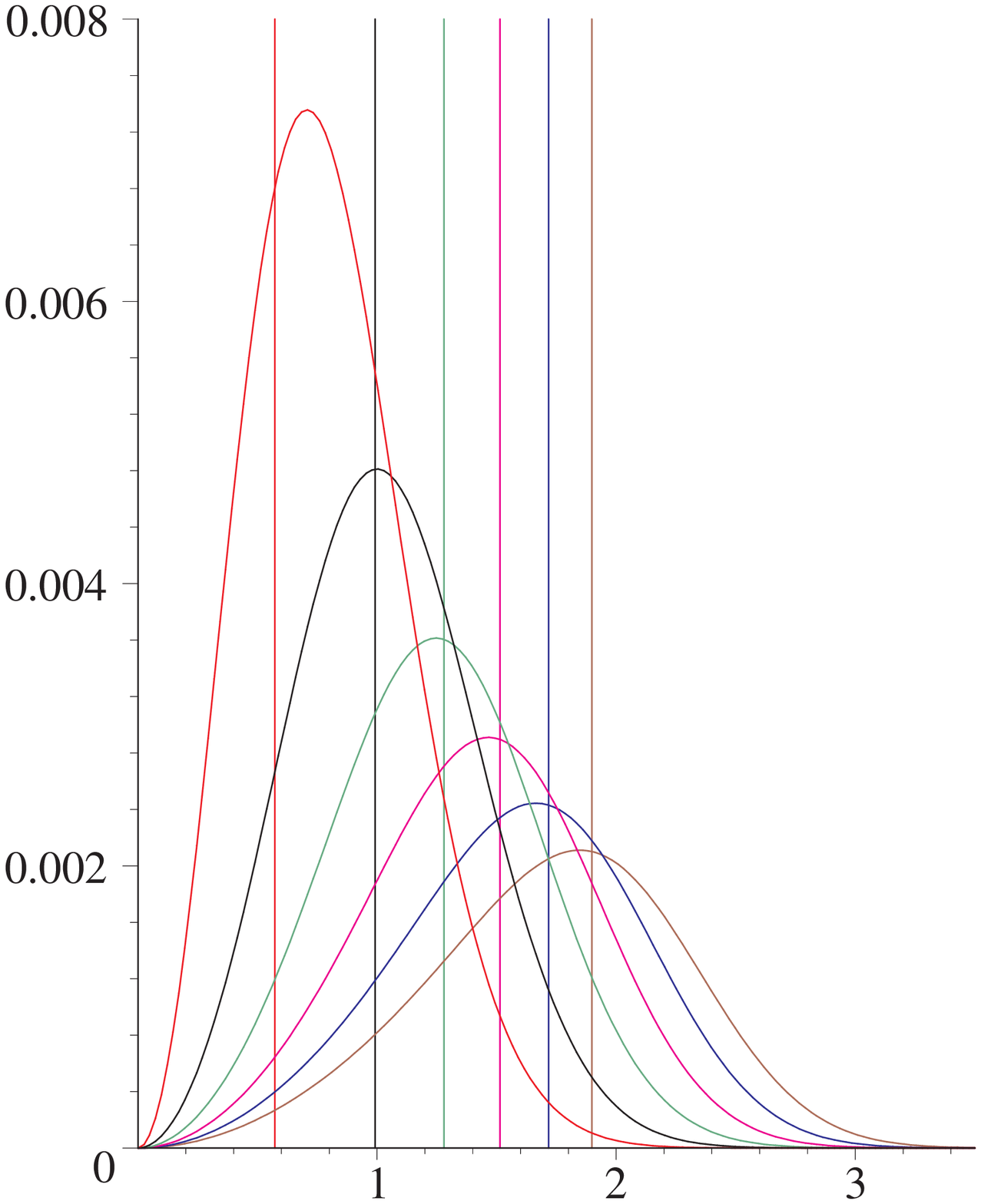}
\put(-210,160){\rotatebox{0}{\mbox{$I$}}}\put(-30,-10){$x[w_0]$}
\caption{Left: Transverse electric field profile,~$E_{P}(x,0,0)$,
and
  Right: transverse intensity profiles~$I_P(x,0,0)$ at focal
  cross-section of Laguerre-Gauss beams comprising superpositions
  $\Psi_P$, $(P=1,...,6)$, according to coefficient
  matrix~(\ref{eq_coeff_matrix}) ($x$-axis in units of
  focal beam radius~$w_{0}$, total cross-sectional beam power
  normalized to unity,
  (${\epsilon_0 \omega_L^2/2}$ set to unity), Rayleigh lengths~$z_R$
  kept constant). The vertical bars mark in the plot on the right are
  located at positions $0.57 \cdot\sqrt{2P+1}\cdot w_{0x}$
  confirming harmonic oscillator-scaling~\cite{Ole_2005AmJPh} of
  the superposition beams' widths.} \label{fig_SuperPos_E&I}
\end{figure}

Mode-superpositions extend the ``useful'' linear part of the field
profile yielding wider parabolic intensity profiles.
Figures~\ref{fig_SuperPos_E&I} and~\ref{fig_PowerCompensate}
demonstrate that the parabolic part in the focal intensity profile
of a superpositions field grows with the number of modes used.

\subsection{Increased Beam Powers Compensate for Potentials'
Widening}\label{SS_Power_Compensation}

\begin{figure}[ht]
\centering
\includegraphics[width=3.0in,height=2.2in]{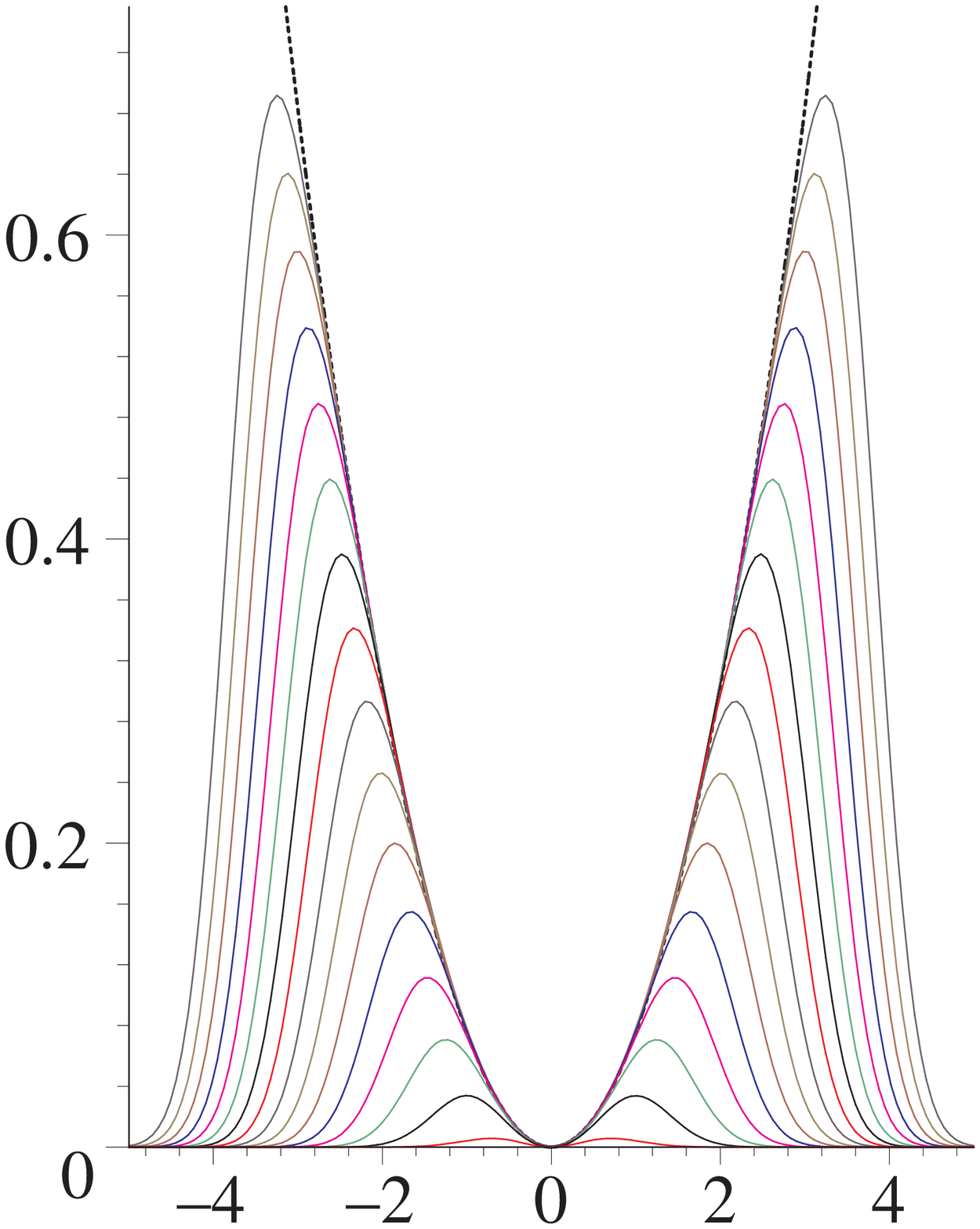}
\put(-210,140){\rotatebox{0}{\mbox{$ I$}}} \put(-30,-10){$x[w_0]$}
\put(-130,70){\rotatebox{0}{\mbox{\colorbox{white}{
     \includegraphics[width=1.7in,height=1.3in]{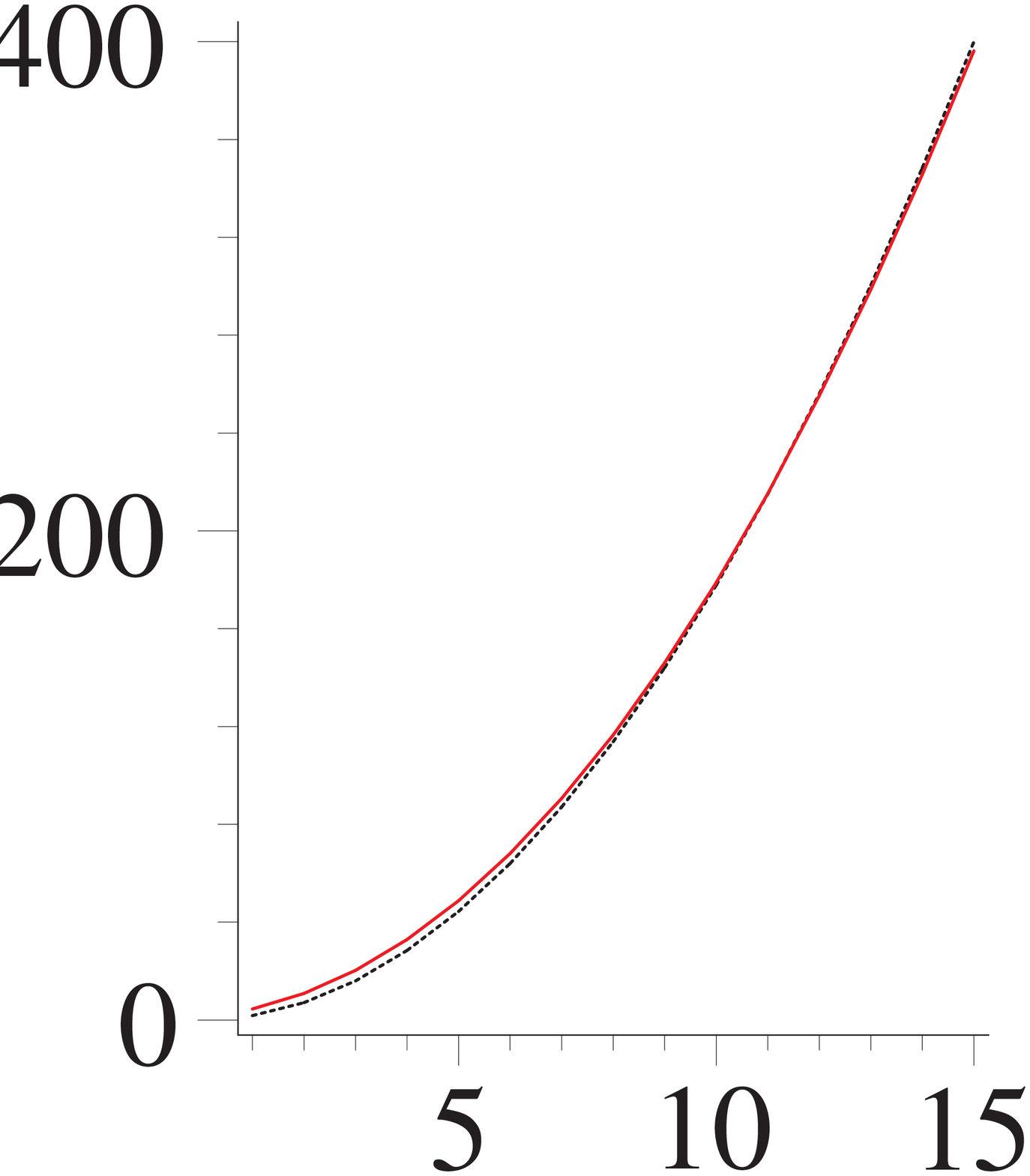}}}}}
\hspace{0.5cm}
\caption{Focal intensity profiles~$I(x,0)$ of
  Laguerre-Gauss beam superpositions
  comprising up to 16${}^{\mbox{th}}$ order modes
  (same units as in Fig.~\ref{fig_SuperPos_E&I}). For the superposition
  modes depicted in the left panel the total beam power has
  been adjusted such that all profiles have the same curvature at the
  origin as the dotted line parabola. The inset shows the necessary
  relative power increase, $\bar I_P/\bar I_1$,
  as a function of superposition order~$P$~(solid red line); it
  scales approximately like $\frac{16}{9}\cdot
  P^{2}$ (inset: dotted black line).} \label{fig_PowerCompensate}
\end{figure}

\begin{figure}
\centering
\includegraphics[width=3.0in,height=2.2in]{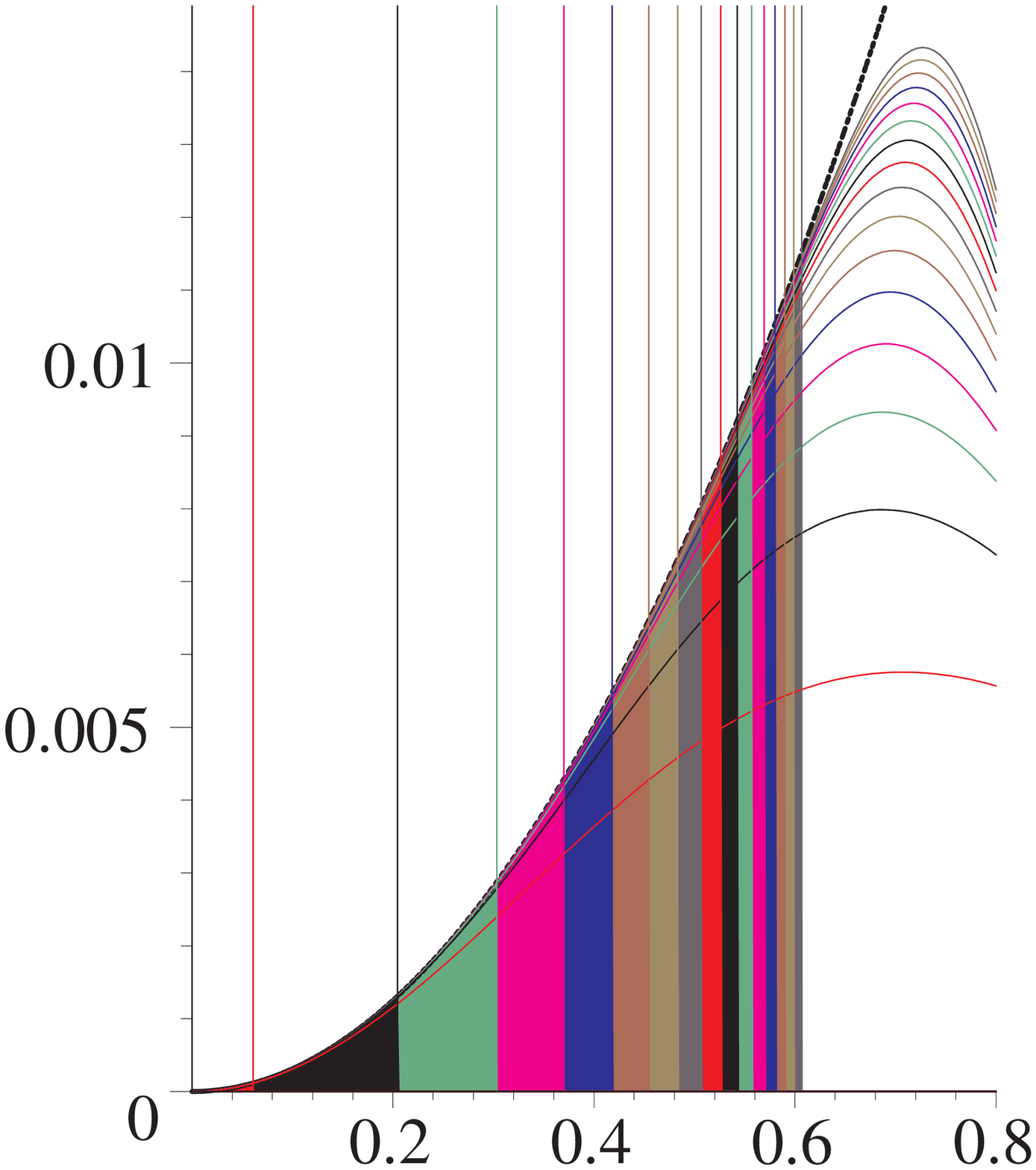}
\put(-210,140){\rotatebox{0}{\mbox{$I$}}} \put(-30,-10){$x[w_0]$}
\caption{Focal intensity profiles~$I(x,0,0)$ of
  Laguerre-Gauss beams comprising up to 16${}^{\mbox{th}}$ order modes
  and their 0.74\%-deviation marks $d_{P}$,
  which lie at relative positions $ d_{P}/d_1 = 1.00, 3.36,\ldots,
  9.96$ from the origin, compare Table~\ref{table:energy} (same units
  as in Fig.~\ref{fig_SuperPos_E&I}). In contrast to Fig.~\ref{fig_PowerCompensate}
  all superpositions have the same total beam power $\bar I$, but the
  Rayleigh lengths~$z_{R}$ have been readjusted such that all
  higher-order superpositions match up with curvature of the first
  mode case~$\Psi_1$, see text.} \label{fig_RayleighLength_Compensate}
\end{figure}

If we increase the total cross-sectional beam power $\bar I_{P}$ for
wider beam profiles according to the ratios of the modes' transverse
derivatives, $\bar I_{P} \doteq \bar I_1 |{\partial_x
\Psi_1(x,0,0)}/ {\partial_x \Psi_{P}(x,0,0)}|^2$, the weakened
gradient is power-compensated for by increased laser power. This way
all superpositions give rise to potentials with equal strength, see
Fig.~\ref{fig_PowerCompensate}, the necessary beam power increase to
achieve this compensation is sketched in the inset of
Fig.~\ref{fig_PowerCompensate}.

\subsection{Decreased Rayleigh-Lengths Compensate for Potentials' Widening}\label{SS_Rayleigh_Length_Matching}

Alternatively to the beam-power increases just discussed, we can
keep the total beam power for all beams equal and shrink the
higher-order superposition-beams' Rayleigh lengths through increased
beam focussing in the $\rho$-direction. This Rayleigh
length-matching also allows us to compensate for the gradient
reduction observed in Fig.~\ref{fig_SuperPos_E&I}. The laser
intensity profiles for Rayleigh-matched superpositions are displayed
in Fig.~\ref{fig_RayleighLength_Compensate}, the filled-in areas in
this figure are limited by the points~$d_{P}$, where each intensity
curve deviates from the enveloping parabola (dotted line) by 0.74
percent. They delineate the useful areas of the potentials. This
quality-criterion is adopted from Gallatin and Gould's
work~\cite{Gallatin_1991OSAJB} which showed that beyond a deviation
of 0.74\% spherical aberrations distort the atomic point-spread
function of an imaged atomic beam too severely; for more details see
references~\cite{Ole_PRA09} and~\cite{Gallatin_1991OSAJB}.

\subsection{Power Savings}\label{SS_PowerSavings}

The filled-in areas in Fig.~\ref{fig_RayleighLength_Compensate}
represent the laser power fraction contributing to the atom
potential in each case. Higher-order superpositions clearly allow us
to use the laser power much more efficiently. Most of the laser
power is wasted in the wings if the superposition approach is not
employed. Additionally to the quantification of the useful area of
the potentials delineated by the deviation points~$d_{P}$ (see
Fig.~\ref{fig_RayleighLength_Compensate} and
Table~\ref{table:energy}). This waste is meaningfully quantified
through the determination of the fraction of power~${\cal E}_{P}$
the laser beam contributes to the `useful' part of the potential
profile.  We define it as the ratio of the laser energy contributing
to the area between the deviation points~$|\rho|<d_{P}$, in terms of
the total laser power, namely

\begin{table}[t]
\caption{Potential Parameters ${d_{P}}$ and ${\cal E}_{P}$, compare
Fig.\ref{fig_RayleighLength_Compensate}}
\centering          
\begin{tabular}{c c c c c c c c c c c c c c c c c c}
\hline\hline                        
$P$ & 1 & 2 & 3 & 4 & 5 & 6 & 7 & 8 & 9 & 10 & 11 & 12
& 13 & 14 & 15 & 16 
 \\      
\hline                      
${d_{P}}/{d_1}$ & 1.00 & 3.36 & 4.98 & 6.07 & 6.86 & 7.46 & 7.93 &
8.31 & 8.63 & 8.90 & 9.14 & 9.34 & 9.52 & 9.68 & 9.83 & 9.96 
 \\
${\cal E}_{P}$ [\%] \; \; & 0.0027 & 0.35 & 1.7 & 3.7 & 6.1 & 8.5 &
11 & 13 & 15 & 17 & 19 & 21 & 23 & 24 & 26 & 27 
 \\
${{\cal E}_{P}}/{{\cal E}_1}$ & 1 & 127 & 614 & 1362 & 2221 & 3103 &
3967 & 4791 & 5570 & 6306 & 6991 & 7639 & 8244 & 8816 & 9352 & 9858
 \\[1ex]
\hline          
\end{tabular}
\label{table:energy}    
\normalsize 
\end{table}

\begin{eqnarray}
{\cal E}_{P} = \frac{ \int_{0}^{d_{P}} d\rho \int_{0}^{2\pi} d\tau
\; I_{P}(\rho \cos(\tau),\rho \sin(\tau),0)}{\int_{-\infty}^{\infty}
dy \int_{-\infty}^{\infty} dx \; I_{P}(x,y,0)} \, .
\label{eq_Rel_power}
\end{eqnarray}

\begin{figure}[h]
\centering
\includegraphics[width=3.in,height=2.5in]{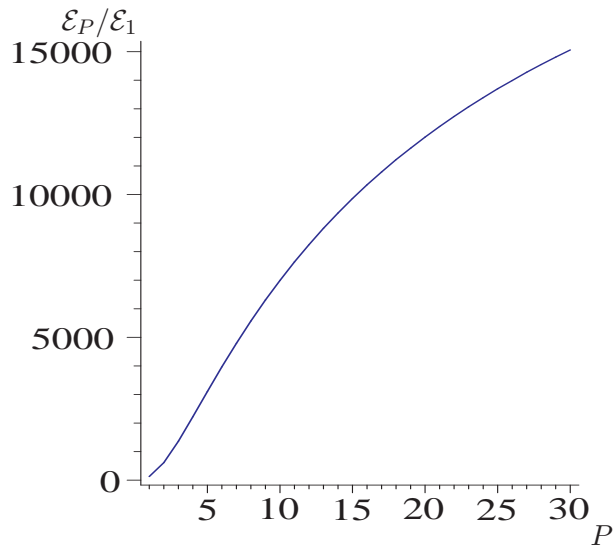}
\put(-200,185){\rotatebox{0}{\mbox{${\cal E}_{P}/{\cal E}_{1}$}}}
\put(-0,-10){$P$}
\caption{Relative power savings~${\cal E}_{P}/{\cal E}_{1}$ as a
function of superposition mode number $P$.} \label{fig_PowerVsModes}
\end{figure}

Table~\ref{table:energy} and Fig.~\ref{fig_PowerVsModes} summarize
and quantify our findings: Table~\ref{table:energy} allows us to
compare values for a single-mode atom potential, for which ${\cal
E}_1 =0.0027\%$, with the superposition approach. For example,
compared to mode~$\Psi_1=LG_{0,1}$ the relative power savings in
case of superposition ${\Psi}_{16}$ is 9858, this translates into a
power utilization of ${\cal E}_{16}=0.0027\% \times 9858 = 27 \%$.
In general the details of this behaviour depend on the chosen
quality criterion but the underlying scaling is straightforward to
derive. The useful fraction of the laser beam is proportional to a
2D integral over the intensity and therefore grows with the fourth
power of the position of the deviation mark ${\cal E}_{P}/{\cal
E}_{1} = (d_{P}/d_1)^4$, for example~${\cal E}_{16}/{\cal E}_{1} =
(d_{16}/d_1)^4 \approx 9.96^4 \approx 9858$.

\subsection{Limitations due to Mode Dispersion}\label{SS_CYLINDRICAL_Mode_Dispersion}

\begin{figure}[h]
\centering
\includegraphics[width=3.in,height=2.5in]{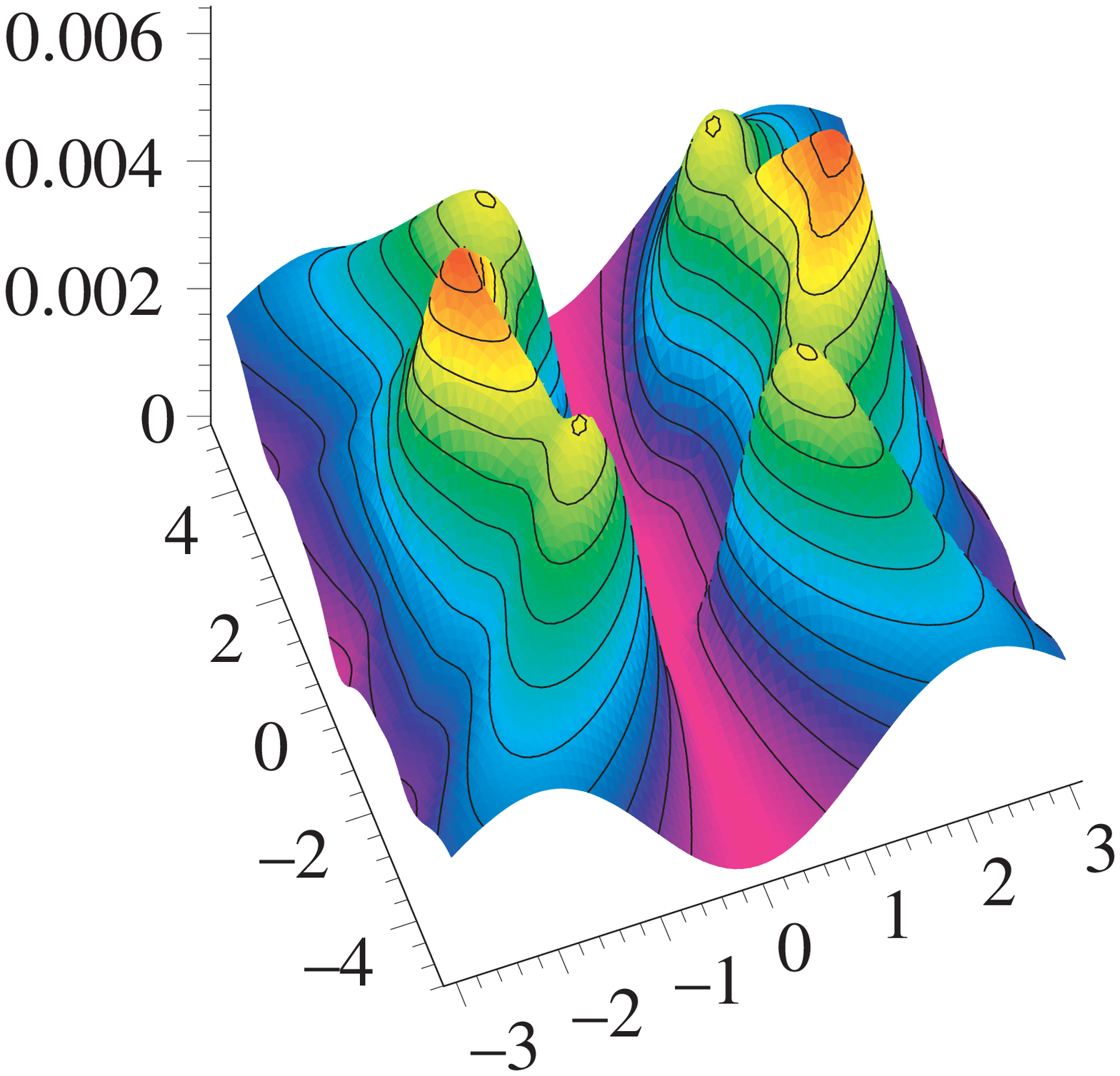}
\put(-200,185){\rotatebox{0}{\mbox{$I$}}} \put(-50,10){$x[w_0]$}
\put(-200,50){$z[z_R]$} \hspace{0.5cm}
\includegraphics[width=3.in,height=2.5in]{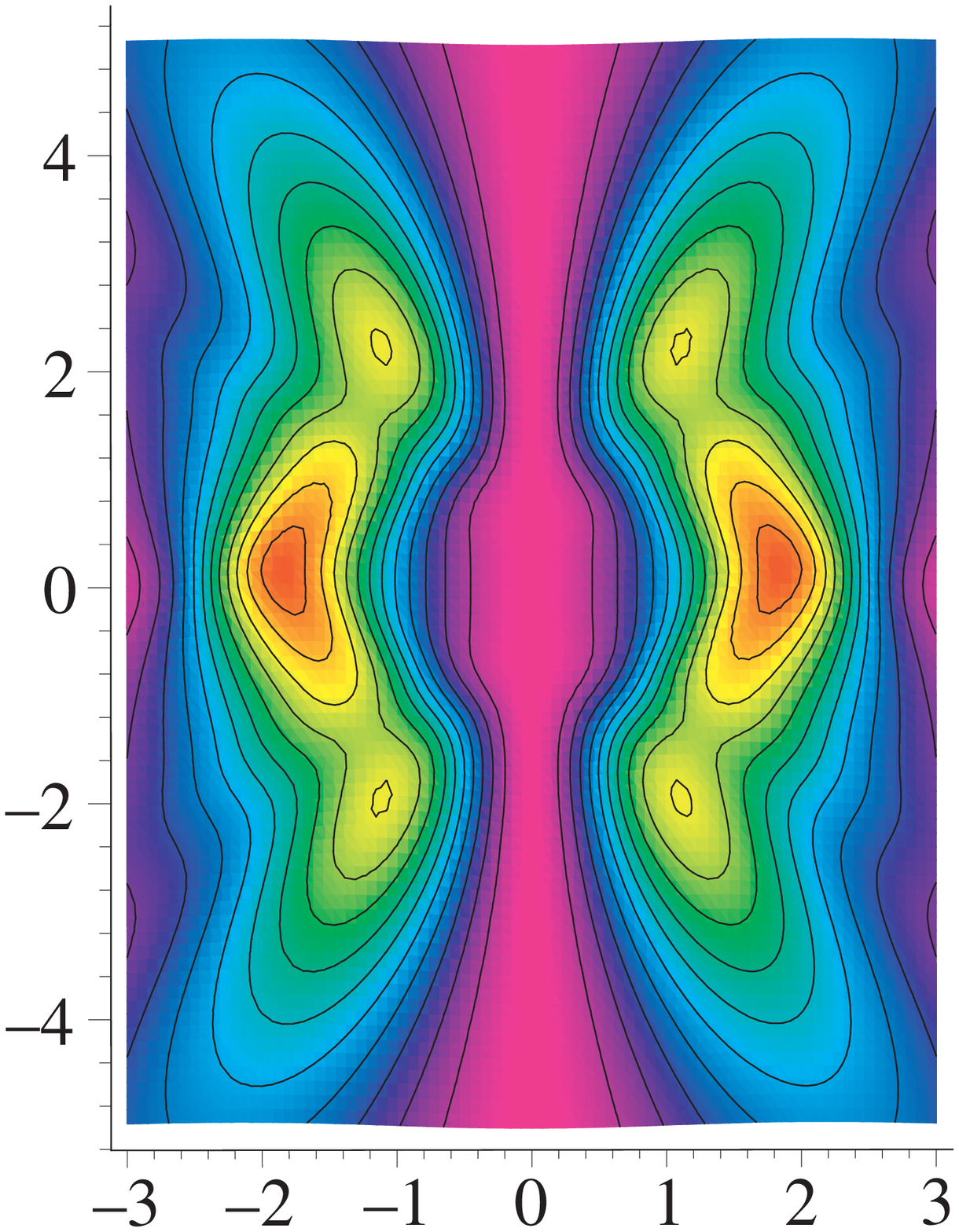}
\put(-220,180){\rotatebox{0}{\mbox{$x[w_0]$}}}\put(-30,-10){$z[z_R]$}
\caption{Intensity profile~$I_{6}(x,0,z)$ of Laguerre-Gauss
superposition~$\Psi_6$, from two perspectives. The mode dispersive
effects of Gouy's phase change the intensity distribution along the
beam axis. Note that the $x$-axis is displayed in terms of the focal
beam radius~$w_0$ whereas the $z$-axis is represented in terms of
the Rayleigh length~$z_R$.} \label{fig_LaguerreGauss_Intensities}
\end{figure}

Different order modes carry different Gouy-phase factors which leads
to mode dispersion in the focal region~\cite{Ole_2005AmJPh}. The
associated change in intensity distribution is illustrated in
Fig.~\ref{fig_LaguerreGauss_Intensities}. An interesting case is the
creation of spherically symmetrical potentials using two beams with
cylindrical symmetry and equal foci orthogonally crossing each
other. In this case the width of one beam dictates the area along the
beam axis of the other beam that is used. This implies constraints on
the degrees of permissible focussing which we investigate in the
following section.

\section{Spherical Potentials \label{S_spherical_Potentials}}

We now want to investigate the scenario for the generation of
spherically symmetrical potentials. If an identical copy of the
laser beam that travels along the $z$-axis is additionally sent
along the $x$-axis such that their crossed confocal configuration
leads to the simultaneous application of two cylindrical potentials
a spherical potential is applied to the gas cloud. The laser beams
have to be sufficiently detuned from each other in order to avoid
harmful interference despite their spatial
overlap~\cite{DePue_AntiJitter_PRL99}. They moreover have to be
elliptically stretched in the $y$-direction, by a factor~$\sqrt{2}$
in the parabolic case or by $2^{1/(2l)}$ in the case of a monomial
potential~$\propto \rho^{2l}$, otherwise adding up their intensities
would remove the desired isotropy of the spherical potential.

\subsection{Limitations due to Mode Dispersion}\label{SS_spherical_Mode_Dispersion}

\begin{figure}[ht]
\begin{center}
  \begin{minipage}[b]{0.28\linewidth} 
     \includegraphics[width=1\linewidth,height=1\linewidth]{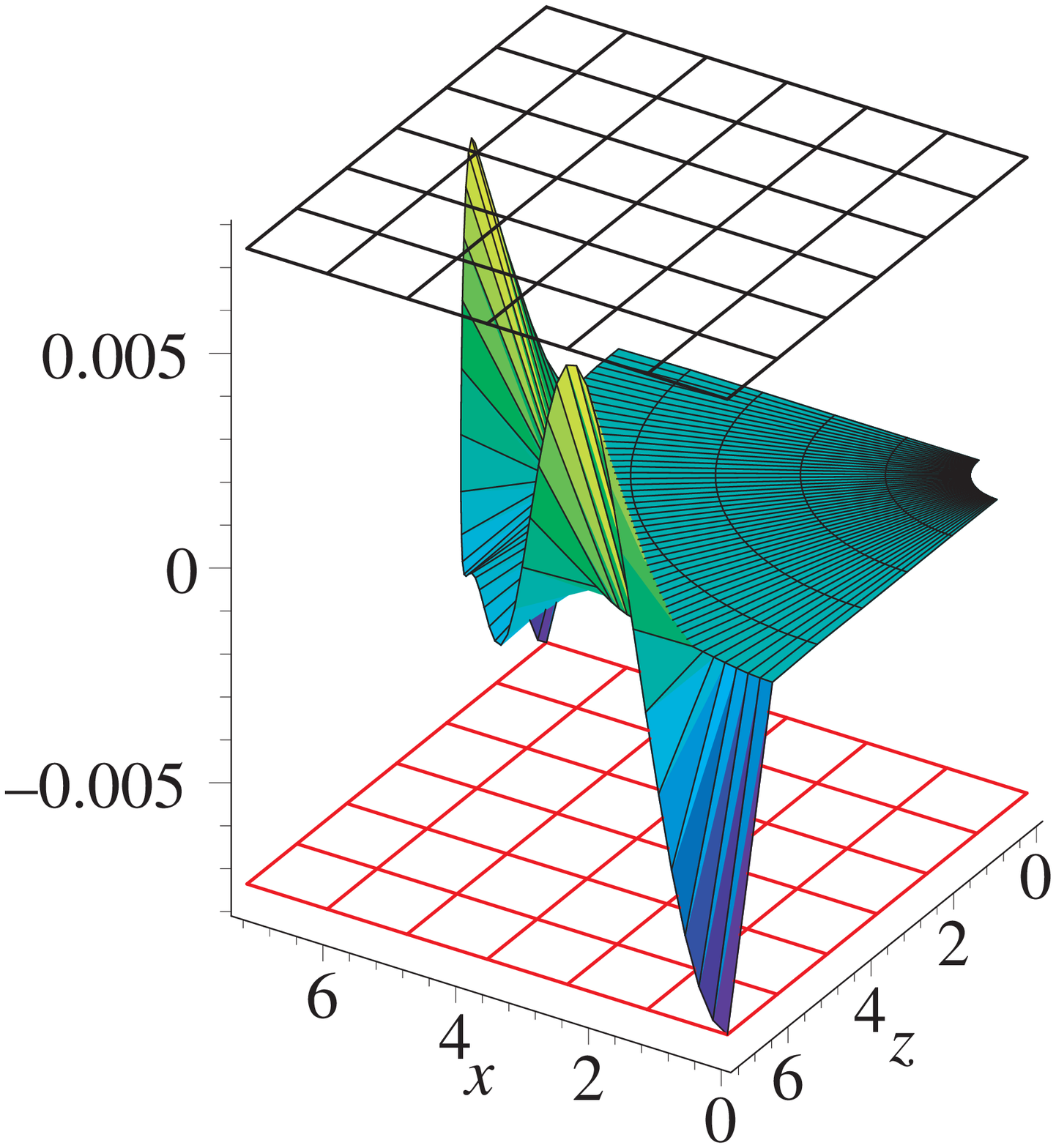}
     \put(-130,110){\rotatebox{0}{\mbox{$\Delta I_{30}$}}} 
  \end{minipage}   
  \hspace{0.045\linewidth}
  \begin{minipage}[b]{0.28\linewidth}
     \includegraphics[width=1\linewidth,height=1\linewidth]{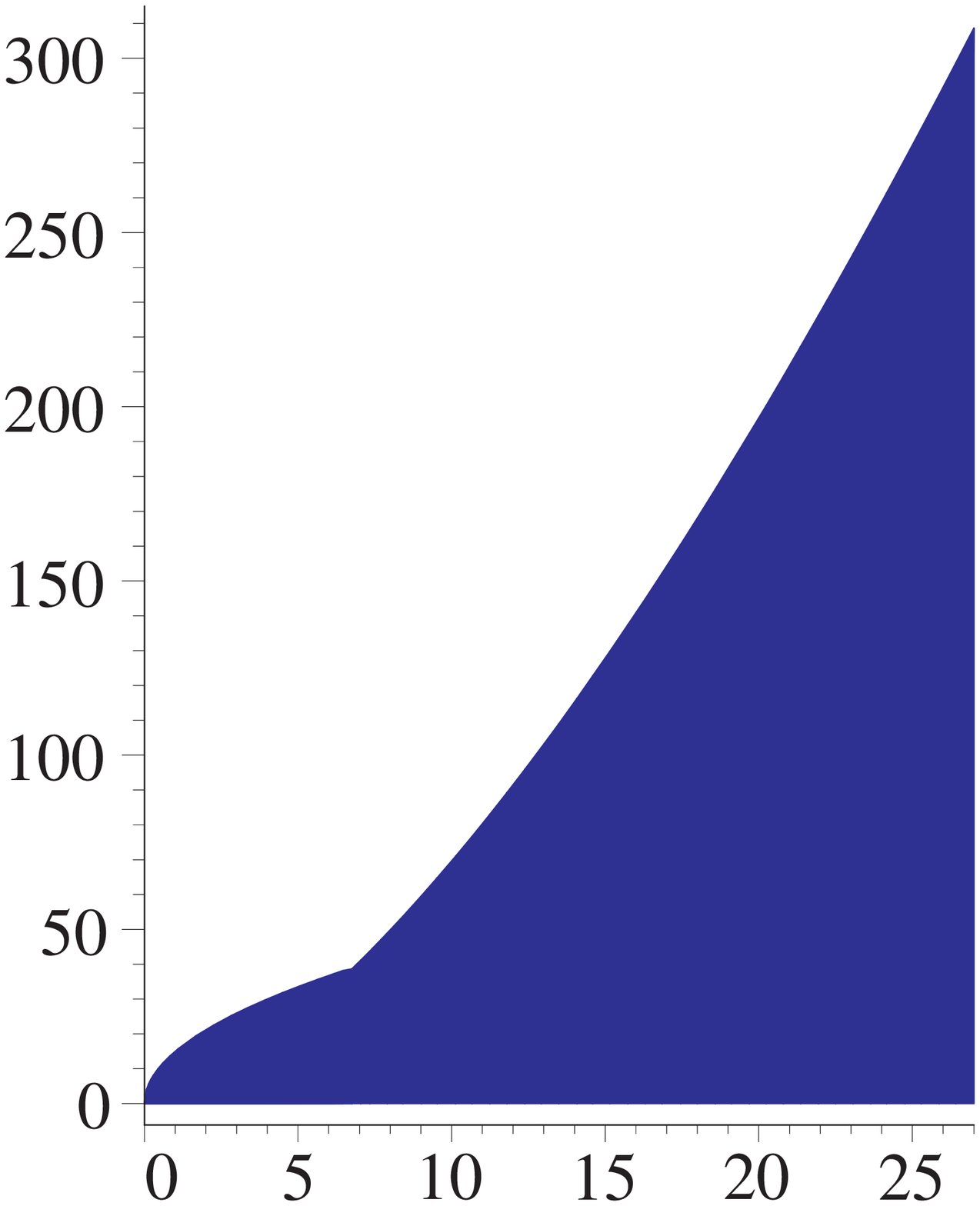}
     \put(-145,50){\rotatebox{90}{\mbox{$z_{\min}(P)/\lambda_L$}}} \put(-30,-10){$P$}
  \end{minipage}
  \hspace{0.045\linewidth}
  \begin{minipage}[b]{0.28\linewidth}
     \includegraphics[width=1\linewidth,height=1\linewidth]{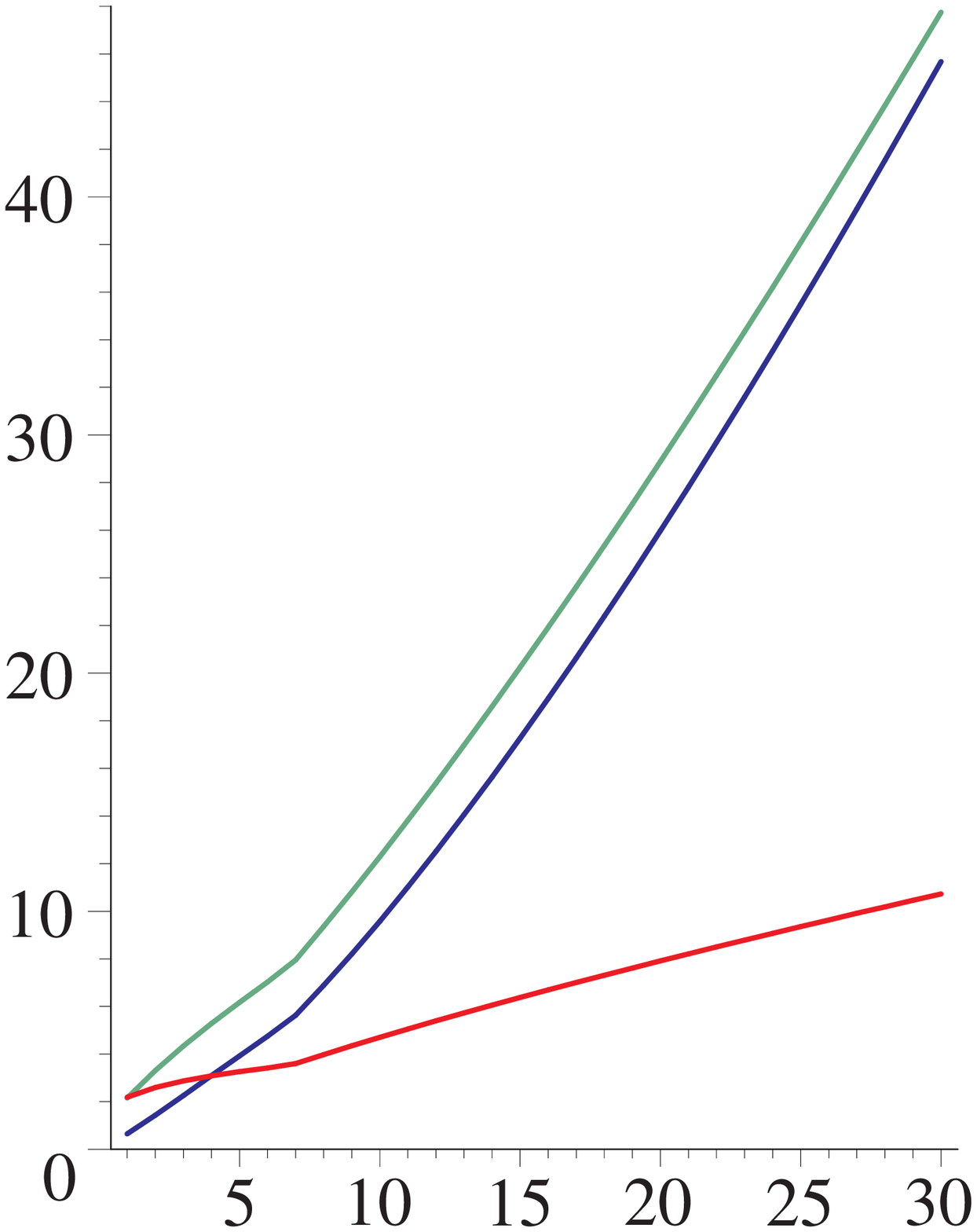}
     \put(-137,120){\rotatebox{0}{\mbox{$[\lambda_L]$}}}
     \put(-30,-10){$P$}
  \end{minipage}
\end{center}
\caption{The left panel illustrates the behavior of the relative
  deviation of the intensity distribution $\Delta I$ from zero as
  it approaches the 0.74\%-deviation marks (top and bottom
  grid). Here, $\Delta I_{30}$ is shown for the crossed
  configuration of two laser beams travelling along $z$ and $x$-axis
  respectively. The value of the Rayleigh length $z_{R}$ at which we
  find that the oscillatory behaviour of $\Delta I$ along a
  constant radial perimeter just exhausts the upper and lower limits
  set by the deviation marks allows us to determine the associated
  value of $z_{\min}$. The latter is plotted as a function of maximum
  mode number, in the middle panel (the filled in blue area is the
  forbidden area of too tightly focussed beams). The values of
  $z_{\min}(P)$ in turn determine the
  position of the turning points $0.57\cdot
  \sqrt{2P+1}\cdot w_0$ (top green line, compare Fig.~\ref{fig_SuperPos_E&I}),
  the position of the deviation-points~$d_{P}$ (middle blue line),
  and the focal beam radii $w_0(z_{min}(P))$ (lower red line), depicted
  in terms of the laser's wavelength $\lambda_L$ in the right panel.}
\label{Fig_Iradial_zMin_alphaMax}
\end{figure}

Gouy's phase~$\phi(z)=\arctan(z/z_R)\approx z /z_R,$ varies
strongest near the beam focus and introduces relative phases between
the modes within each beam~\cite{Ole_2005AmJPh}. If the beam is very
strongly focussed (small value of~$z_{R}$) the dephasing away from
the focus $z=0$ is so rapid that non-linear aberrations degrade the
desired linear field profile. In other words, a lower limit for the
Rayleigh lengths~$z_{\min}(P)$ as a function of the number of used
modes~$P$ has to be determined in order to guarantee moderate
dephasing. Whereas the absolute values for this lower limit are hard
to derive from first principles, we can still work out the correct
scaling with the maximal mode number~$P$:

The electric field is proportional to the superposition of the modes
including the Gouy-phase factors; this can be approximated by $E_{P}
\propto \sum_{p=0}^{P-1} c_{p+1} LG_{p} e^{2 i p \phi} \approx
\sum_{p=0}^{P-1} c_{p+1} LG_{p} ( 1 + 2 i p z/z_{R}) $. The
expansion coefficients are positive and the wave functions are real
at the focus $z=0$. Since the first order term is purely imaginary
the intensity has to depend on $z$ quadratically: $I_{P}(x,y,z) =
I_{P}(x,y,0) \cdot [1+\frac{z^2}{z^2_{R}} D_{P} + {\cal O}(z^4)]$.
The deviation term $D_{P}$ has a complicated dependence on the
number of modes, but, containing the square of sums of the
form~$\sum_{p=0}^{P-1} p c_{p+1} LG_{p} $, is roughly proportional
to ${{P}}^2$. When we consider the relative deviation of the
intensity profile near the focus from the focal intensity
distribution, $\Delta I = \frac{I(z)-I(0)}{I(0)}$, we find $\Delta
I_{P} \propto \frac{z^2}{z^2_{R}} \cdot {{P}}^2 $. Additionally, we
know that the widths of the superpositions scale roughly like those
of the harmonic oscillator~\cite{Ole_2005AmJPh}, see
Fig.~\ref{fig_SuperPos_E&I}, namely $z\propto \sqrt{2P+1}\approx
\sqrt{2} \sqrt{P}$. For constant relative intensity deviations
$\Delta I_{P}$ this implies $const. = \frac{\sqrt{P}^2}{z^2_{R}}
\cdot P^2$ or $z_{R}\propto {{P}}^{3/2}$. A numerical investigation,
see Fig.~\ref{Fig_Iradial_zMin_alphaMax}, confirms $z_{\min}(P) =
2.2 \cdot \lambda_L \cdot P^{3/2}$ as a good estimate for a lower
bound on~$z_{R}$. This relationship has been checked numerically and
holds for $7 < P < 30$. There is no reason to believe deviations
might occur for values of $ P > 30$, but for small values of $P$ the
assumptions used in the derivation of the scaling law do not hold
accurately, see Fig.~\ref{fig_SuperPos_E&I}. Instead, the expression
$z_{\min}{{P}} = 15 \cdot \lambda_L \cdot P^{1/2}$ gives a better
estimate for $z_{\min}{{P}}$ in the range of $0 < P \leq 6$. These
lower limits for $z_{R}$ imply that the beam focus is several
wavelengths wide and {a posteriori} confirms that the paraxial
approximations hold for all cases discussed here, since the largest
beam opening angle conforming with the lower limits presented here
turns out to be roughly $7^\circ$ for superposition~$\Psi_2$.

\section{Conclusions}\label{S_Conclusion}

For a possible experimental implementation of focussing fields
repulsive (blue-detuned) optical potentials with a dark center are
probably most suited since they minimize detrimental spontaneous
emission noise. For an expansion field analogously red detuned
optical potentials~\cite{Jeltes_NAT07} should be used. If
fine-tuning is considered one will probably also have to revisit the
approximations underlying~Eq.~(\ref{Intensity}) and
Eq.~(\ref{dipole_potential_approx}); such considerations are beyond
the scope of this paper.

The techniques for the coherent superposition of laser modes have
been experimentally demonstrated, see e.g.
references~\cite{OleJMO05,Maurer_RitschMarte_NJP07} and citations
therein. We have found here that using the mode-superposition
approach allows for very considerable laser power savings and
potentials can be made wider than is possible with pure modes. We
come to the conclusion that for the design of atomic potentials,
based on the optical dipole force, it is possible and necessary to
coherently superpose suitable laser modes in order to create wide
high quality parabolic potentials.

\bibliography{OpticalAtomPotential_3D}

\end{document}